\documentstyle[preprint,aps]{revtex}
\begin{document}
\title{{\bf Relativistic treatment in }$D$-{\bf dimensions to a spin-zero particle
with noncentral equal scalar and vector ring-shaped Kratzer potential }}
\author{Sameer M. Ikhdair\thanks{%
sikhdair@neu.edu.tr} and \ Ramazan Sever\thanks{%
sever@metu.edu.tr}}
\address{$^{\ast }$Department of Physics, \ Near East University, Nicosia, North
Cyprus, Mersin 10, Turkey\\
$^{\dagger }$Department of Physics, Middle East Technical University, 06531
Ankara, Turkey.}
\date{\today
}
\maketitle

\begin{abstract}
The Klein-Gordon equation in $D$-dimensions for a recently proposed Kratzer
potential plus ring-shaped potential is solved analytically by means of the
conventional Nikiforov-Uvarov method. The exact energy bound-states and the
corresponding wave functions of the Klein-Gordon are obtained in the
presence of the noncentral equal scalar and vector potentials. The results
obtained in this work are more general and can be reduced to the standard
forms in three-dimensions given by other works.


\end{abstract}

% INITIALIZE - DONT CHANGE % %  %

\section{Introduction}

\noindent In various physical applications including those in nuclear
physics and high energy physics [1,2], one of the interesting problems is to
obtain exact solutions of the relativistic equations like Klein-Gordon and
Dirac equations for mixed vector and scalar potential. The Klein-Gordon and
Dirac wave equations are frequently used to describe the particle dynamics
in relativistic quantum mechanics. The Klein-Gordon equation has also been
used to understand the motion of a spin-$0$ particle in large class of
potentials. In recent years, much efforts have been paid to solve these
relativistic wave equations for various potentials by using different
methods. These relativistic equations contain two objects: the four-vector
linear momentum operator and the scalar rest mass. They allow us to
introduce two types of potential coupling, which are the four-vector
potential (V) and the space-time scalar potential (S).

Recently, many authors have worked on solving these equations with physical
potentials including Morse potential [3], Hulthen potential [4], Woods-Saxon
potential [5], P\"{o}sch-Teller potential [6], reflectionless-type potential
[7], pseudoharmonic oscillator [8], ring-shaped harmonic oscillator [9], $%
V_{0}\tanh ^{2}(r/r_{0})$ potential [10], five-parameter exponential
potential [11], Rosen-Morse potential [12], and generalized symmetrical
double-well potential [13], etc. It is remarkable that in most works in this
area, the scalar and vector potentials are almost taken to be equal (i.e., $%
S=V$) [2,14]. However, in some few other cases, it is considered the case
where the scalar potential is greater than the vector potential (in order to
guarantee the existence of Klein-Gordon bound states) (i.e., $S>V$) [15-19].
Nonetheless, such physical potentials are very few. The bound-state
solutions for the last case is obtained for the exponential potential for
the $s$-wave Klein-Gordon equation when the scalar potential is greater than
the vector potential [15].

The study of exact solutions of the nonrelativistic equation for a class of
non-central potentials with a vector potential and a non-central scalar
potential is of considerable interest in quantum chemistry [20-22]. In
recent years, numerous studies [23] have been made in analyzing the bound
states of an electron in a Coulomb field with simultaneous presence of
Aharanov-Bohm (AB) [24] field, and/or a magnetic Dirac monopole [25], and
Aharanov-Bohm plus oscillator (ABO) systems. In most of these works, the
eigenvalues and eigenfunctions are obtained by means of seperation of
variables in spherical or other orthogonal curvilinear coordinate systems.
The path integral for particles moving in non-central potentials is
evaluated to derive the energy spectrum of this system analytically [26]. In
addition, the idea of SUSY and shape invariance is also used to obtain exact
solutions of such non-central but seperable potentials [27,28]. Very
recently, the conventional Nikiforov-Uvarov (NU) method [29] has been used
to give a clear recipe of how to obtain an explicit exact bound-states
solutions for the energy eigenvalues and their corresponding wave functions
in terms of orthogonal polynomials for a class of non-central potentials
[30].

Another type of noncentral potentials is the ring-shaped Kratzer potential,
which is a combination of a Coulomb potential plus an inverse square
potential plus a noncentral angular part [31,32]. The Kratzer potential has
been used to describe the vibrational-rotational motion of isolated diatomic
molecules [33] and has a mixed-energy spectrum containing both bound and
scattering states with bound-states have been widely used in molecular
spectroscopy [34]. The ring-shaped Kratzer potential consists of radial and
angular-dependent potentials and is useful in studying ring-shaped molecules
[22]. In taking the relativistic effects into account for spin-$0$ particle
in the presence of a class of noncentral potentials, Yasuk {\it et al} [35]
applied the NU method to solve the Klein-Gordon equation for the noncentral
Coulombic ring-shaped potential [21] for the case $V=S.$ Further, Berkdemir
[36] also used the same method to solve the Klein-Gordon equation for the
Kratzer-type potential.

Recently, Chen and Dong [37] proposed a new ring-shaped potential and
obtained the exact solution of the Schr\"{o}dinger equation for the Coulomb
potential plus this new ring-shaped potential which has possible
applications to ring-shaped organic molecules like cyclic polyenes and
benzene. This type of potential used by Chen and Dong [37] appears to be
very similar to the potential used by Yasuk {\it et al} [35]. Moreover,
Cheng and Dai [38], proposed a new potential consisting from the modified
Kratzer's potential [33] plus the new proposed ring-shaped potential in
[37]. They have presented the energy eigenvalues for this proposed
exactly-solvable non-central potential in three dimensional $($i.e., $D=3)$%
-Schr\"{o}dinger equation by means of the NU method. The two quantum systems
solved by Refs [37,38] are closely relevant to each other as they deal with
a Coulombic field interaction except for a slight change in the angular
momentum barrier acts as a repulsive core which is for any arbitrary angular
momentum $\ell $ prevents collapse of the system in any dimensional space
due to the slight perturbation to the original angular momentum barrier.
Very recently, we have also applied the NU method to solve the
Schr\"{o}dinger equation in any arbitrary $D$-dimension to this new modified
Kratzer-type potential [39,40].

The aim of the present paper is to consider the relativistic effects for the
spin-$0$ particle in our recent works [39,40]. So we want to present a
systematic recipe to solving the $D$-dimensional Klein-Gordon equation for
the Kratzer plus the new ring-shaped potential proposed in [38] using the
simple NU method. This method is based on solving the Klein-Gordon equation
by reducing it to a generalized hypergeometric equation.

This work is organized as follows: in section \ref{TKG}, we shall present
the Klein-Gordon equation in spherical coordinates for spin-$0$ particle in
the presence of equal scalar and vector noncentral Kratzer plus the new
ring-shaped potential and we also separate it into radial and angular parts.
Section \ref{NUM} is devoted to a brief description of the NU method. In
section \ref{ES}, we present the exact solutions to the radial and angular
parts of the Klein-Gordon equation in $D$-dimensions. Finally, the relevant
conclusions are given in section \ref{C}.

\section{The Klein-Gordon Equation with Equal Scalar and Vector Potentials}

\label{TKG}In relativistic quantum mechanics, we usually use the
Klein-Gordon equation for describing a scalar particle, i.e., the spin-$0$
particle dynamics. The discussion of the relativistic behavior of spin-zero
particles requires understanding the single particle spectrum and the exact
solutions to the Klein Gordon equation which are constructed by using the
four-vector potential ${\bf A}_{\lambda }$ $(\lambda =0,1,2,3)$ and the
scalar potential $(S)$. In order to simplify the solution of the
Klein-Gordon equation, the four-vector potential can be written as ${\bf A}%
_{\lambda }=(A_{0},0,0,0).$ The first component of the four-vector potential
is represented by a vector potential $(V),$ i.e., $A_{0}=V.$ In this case,
the motion of a relativistic spin-$0$ particle in a potential is described
by the Klein-Gordon equation with the potentials $V$ and $S$ [1]$.$ For the
case $S\geq V,$ there exist bound-state (real) solutions for a relativistic
spin-zero particle [15-19]. On the other hand, for $S=V,$ the Klein-Gordon
equation reduces to a Schr\"{o}dinger-like equation and thereby the
bound-state solutions are easily obtained by using the well-known methods
developed in nonrelativistic quantum mechanics [2].

The Klein-Gordon equation describing a scalar particle (spin-$0$ particle)
with scalar $S(r,\theta ,\varphi )$ and vector $V(r,\theta ,\varphi )$
potentials is given by [2,14]

\begin{equation}
\left\{ {\bf P}^{2}-\left[ E_{R}-V(r,\theta ,\varphi )/2\right] ^{2}+\left[
\mu +S(r,\theta ,\varphi )/2\right] ^{2}\right\} \psi (r,\theta ,\varphi )=0,
\end{equation}
where $E_{R},{\bf P}$ and $\mu $ are the relativistic energy, momentum
operator and rest mass of the particle, respectively. The potential terms
are scaled in (1) by Alhaidari {\it et al }[14] so that in the
nonrelativistic limit the interaction potential becomes $V.$

In this work, we consider the equal scalar and vector potentials case, that
is, $S(r,\theta ,\varphi )=V(r,\theta ,\varphi )$ with the recently proposed
general non-central potential taken in the form of the Kratzer plus
ring-shaped potential [38-40]:
\begin{equation}
V(r,\theta ,\varphi )=V_{1}(r)+\frac{V_{2}(\theta )}{r^{2}}+\frac{%
V_{3}(\varphi )}{r^{2}\sin ^{2}\theta },
\end{equation}
\begin{equation}
V_{1}(r)=-\frac{A}{r}+\frac{B}{r^{2}},\text{ }V_{2}(\theta )=Cctg^{2}\theta ,%
\text{ }V_{3}(\varphi )=0,
\end{equation}
where $A=2a_{0}r_{0},$ $B=a_{0}r_{0}^{2}$ and $C$ is positive real constant
with $a_{0}$ is the dissociation energy and $r_{0}$ is the equilibrium
internuclear distance [33]. The potentials in Eq. (3) introduced by
Cheng-Dai [38] reduce to the Kratzer potential in the limiting case of $C=0$
[33]$.$ In fact the energy spectrum for this potential can be obtained
directly by considering it as special case of the general non-central
seperable potentials [30].

In the relativistic atomic units ($\hbar =c=1$), the $D$-dimensional
Klein-Gordon equation in (1) becomes [41]

\[
\left\{ \frac{1}{r^{D-1}}\frac{\partial }{\partial r}\left( r^{D-1}\frac{%
\partial }{\partial r}\right) +\frac{1}{r^{2}}\left[ \frac{1}{\sin \theta }%
\frac{\partial }{\partial \theta }\left( \sin \theta \frac{\partial }{%
\partial \theta }\right) +\frac{1}{\sin ^{2}\theta }\frac{\partial ^{2}}{%
\partial \varphi ^{2}}\right] \right.
\]

\begin{equation}
-\left. \left( E_{R}+\mu \right) \left( V_{1}(r)+\frac{V_{2}(\theta )}{r^{2}}%
+\frac{V_{3}(\varphi )}{r^{2}\sin ^{2}\theta }\right) +\left( E_{R}^{2}-\mu
^{2}\right) \right\} \psi (r,\theta ,\varphi )=0.
\end{equation}
with $\psi (r,\theta ,\varphi )$ being the spherical total wave function
separated as follows

\begin{equation}
\psi _{njm}(r,\theta ,\varphi )=R(r)Y_{j}^{m}(\theta ,\varphi ),\text{ }%
R(r)=r^{-(D-1)/2}g(r),\text{ }Y_{j}^{m}(\theta ,\varphi )=H(\theta )\Phi
(\varphi ).
\end{equation}
Inserting Eqs (3) and (5) into Eq. (4) and using the method of separation of
variables, the following differential equations are obtained:

\begin{equation}
\frac{1}{r^{D-1}}\frac{d}{dr}\left( r^{D-1}\frac{dR(r)}{dr}\right) -\left[
\frac{j(j+D-2)}{r^{2}}+\alpha _{2}^{2}\left( \alpha _{1}^{2}-\frac{A}{r}+%
\frac{B}{r^{2}}\right) \right] R(r)=0,
\end{equation}

\begin{equation}
\left[ \frac{1}{\sin \theta }\frac{d}{d\theta }\left( \sin \theta \frac{d}{%
d\theta }\right) -\frac{m^{2}+C\alpha _{2}^{2}\cos ^{2}\theta }{\sin
^{2}\theta }+j(j+D-2)\right] H(\theta )=0,
\end{equation}
\begin{equation}
\frac{d^{2}\Phi (\varphi )}{d\varphi ^{2}}+m^{2}\Phi (\varphi )=0,
\end{equation}
where $\alpha _{1}^{2}=\mu -E_{R},$ $\alpha _{2}^{2}=\mu +E_{R},$ $m$ and $j$
are constants and with $m^{2}$ and $\lambda _{j}=j(j+D-2)$ are the
separation constants.

For a nonrelativistic treatment with the same potential, the Schr\"{o}dinger
equation in spherical coordinates is

\[
\left\{ \frac{1}{r^{D-1}}\frac{\partial }{\partial r}\left( r^{D-1}\frac{%
\partial }{\partial r}\right) +\frac{1}{r^{2}}\left[ \frac{1}{\sin \theta }%
\frac{\partial }{\partial \theta }\left( \sin \theta \frac{\partial }{%
\partial \theta }\right) +\frac{1}{\sin ^{2}\theta }\frac{\partial ^{2}}{%
\partial \varphi ^{2}}\right] \right.
\]

\begin{equation}
+\left. 2\mu \left[ E_{NR}-V_{1}(r)-\frac{V_{2}(\theta )}{r^{2}}-\frac{%
V_{3}(\varphi )}{r^{2}\sin ^{2}\theta }\right] \right\} \psi (r,\theta
,\varphi )=0.
\end{equation}
where $\mu $ and $E_{NR}$ are the reduced mass and the nonrelativistic
energy, respectively. Besides, the spherical total wave function appearing
in Eq. (9) has the same representation as in Eq. (5) but with the
transformation $j\rightarrow \ell $. Inserting Eq. (5) into Eq. (9) leads to
the following differential equations [39,40]:

\begin{equation}
\frac{1}{r^{D-1}}\frac{d}{dr}\left( r^{D-1}\frac{dR(r)}{dr}\right) -\left[
\frac{\lambda _{D}}{r^{2}}-2\mu \left( E_{NR}+\frac{A}{r}-\frac{B}{r^{2}}%
\right) \right] R(r)=0,
\end{equation}

\begin{equation}
\left[ \frac{1}{\sin \theta }\frac{d}{d\theta }\left( \sin \theta \frac{d}{%
d\theta }\right) -\frac{m^{2}+2\mu C\cos ^{2}\theta }{\sin ^{2}\theta }+\ell
(\ell +D-2)\right] H(\theta )=0,
\end{equation}
\begin{equation}
\frac{d^{2}\Phi (\varphi )}{d\varphi ^{2}}+m^{2}\Phi (\varphi )=0,
\end{equation}
where $m^{2}$ and $\lambda _{\ell }=\ell (\ell +D-2)$ are the separation
constants. Equations (6)-(8) have the same functional form as Eqs (10)-(12).
Therefore, the solution of the Klein-Gordon equation can be reduced to the
solution of the Schr\"{o}dinger equation with the appropriate choice of
parameters: \ $j\rightarrow \ell ,$ $\alpha _{1}^{2}\rightarrow -E_{NR\text{
}}$ and $\alpha _{2}^{2}\rightarrow 2\mu .$

The solution of Eq. (8) is well-known periodic and must satisfy the period
boundary condition $\Phi (\varphi +2\pi )=\Phi (\varphi )$ which is the
azimuthal angle solution:
\begin{equation}
\Phi _{m}(\varphi )=\frac{1}{\sqrt{2\pi }}\exp (\pm im\varphi ),\text{ \ }%
m=0,1,2,.....
\end{equation}

Additionally, Eqs (6) and (7) are radial and polar angle equations and they
will be solved by using the Nikiforov-Uvarov (NU) method [29] which is given
briefly in the following section.

\section{Nikiforov-Uvarov Method}

\label{NUM}The NU method is based on reducing the second-order differential
equation to a generalized equation of hypergeometric type [29]. In this
sense, the Schr\"{o}dinger equation, after employing an appropriate
coordinate transformation $s=s(r),$ transforms to the following form:
\begin{equation}
\psi _{n}^{\prime \prime }(s)+\frac{\widetilde{\tau }(s)}{\sigma (s)}\psi
_{n}^{\prime }(s)+\frac{\widetilde{\sigma }(s)}{\sigma ^{2}(s)}\psi
_{n}(s)=0,
\end{equation}
where $\sigma (s)$ and $\widetilde{\sigma }(s)$ are polynomials, at most of
second-degree, and $\widetilde{\tau }(s)$ is a first-degree polynomial.
Using a wave function, $\psi _{n}(s),$ of \ the simple ansatz:

\begin{equation}
\psi _{n}(s)=\phi _{n}(s)y_{n}(s),
\end{equation}
reduces (14) into an equation of a hypergeometric type

\begin{equation}
\sigma (s)y_{n}^{\prime \prime }(s)+\tau (s)y_{n}^{\prime }(s)+\lambda
y_{n}(s)=0,
\end{equation}
where

\begin{equation}
\sigma (s)=\pi (s)\frac{\phi (s)}{\phi ^{\prime }(s)},
\end{equation}

\begin{equation}
\tau (s)=\widetilde{\tau }(s)+2\pi (s),\text{ }\tau ^{\prime }(s)<0,
\end{equation}
and $\lambda $ is a parameter defined as
\begin{equation}
\lambda =\lambda _{n}=-n\tau ^{\prime }(s)-\frac{n\left( n-1\right) }{2}%
\sigma ^{\prime \prime }(s),\text{ \ \ \ \ \ \ }n=0,1,2,....
\end{equation}
The polynomial $\tau (s)$ with the parameter $s$ and prime factors show the
differentials at first degree be negative. It is worthwhile to note that $%
\lambda $ or $\lambda _{n}$ are obtained from a particular solution of the
form $y(s)=y_{n}(s)$ which is a polynomial of degree $n.$ Further, the other
part $y_{n}(s)$ of the wave function (14) is the hypergeometric-type
function whose polynomial solutions are given by Rodrigues relation

\begin{equation}
y_{n}(s)=\frac{B_{n}}{\rho (s)}\frac{d^{n}}{ds^{n}}\left[ \sigma ^{n}(s)\rho
(s)\right] ,
\end{equation}
where $B_{n}$ is the normalization constant and the weight function $\rho
(s) $ must satisfy the condition [29]

\begin{equation}
\frac{d}{ds}w(s)=\frac{\tau (s)}{\sigma (s)}w(s),\text{ }w(s)=\sigma (s)\rho
(s).
\end{equation}
The function $\pi $ and the parameter $\lambda $ are defined as

\begin{equation}
\pi (s)=\frac{\sigma ^{\prime }(s)-\widetilde{\tau }(s)}{2}\pm \sqrt{\left(
\frac{\sigma ^{\prime }(s)-\widetilde{\tau }(s)}{2}\right) ^{2}-\widetilde{%
\sigma }(s)+k\sigma (s)},
\end{equation}
\begin{equation}
\lambda =k+\pi ^{\prime }(s).
\end{equation}
In principle, since $\pi (s)$ has to be a polynomial of degree at most one,
the expression under the square root sign in (22) can be arranged to be the
square of a polynomial of first degree [29]. This is possible only if its
discriminant is zero. In this case, an equation for $k$ is obtained. After
solving this equation, the obtained values of $k$ are substituted in (22).
In addition, by comparing equations (19) and (23), we obtain the energy
eigenvalues.

\section{Exact Solutions of the Radial and Angle-Dependent Equations}

\label{ES}

\subsection{Separating variables of the Klein-Gordon equation}

We seek to solving the radial and angular parts of \ the Klein-Gordon
equation given by Eqs (6) and (7). Equation (6) involving the radial part
can be written simply in the following form [39-41]:
\begin{equation}
\frac{d^{2}g(r)}{dr^{2}}-\left[ \frac{(M-1)(M-3)}{4r^{2}}-\alpha
_{2}^{2}\left( \frac{A}{r}-\frac{B}{r^{2}}\right) +\alpha _{1}^{2}\alpha
_{2}^{2}\right] g(r)=0,
\end{equation}
where
\begin{equation}
M=D+2j.
\end{equation}
On the other hand, Eq. (7) involving the angular part of Klein-Gordon
equation retakes the simple form
\begin{equation}
\frac{d^{2}H(\theta )}{d\theta ^{2}}+ctg(\theta )\frac{dH(\theta )}{d\theta }%
-\left[ \frac{m^{2}+C\alpha _{2}^{2}\cos ^{2}\theta }{\sin ^{2}\theta }%
-j(j+D-2)\right] H(\theta )=0.
\end{equation}
Thus, Eqs (24) and (26) have to be solved latter through the NU method in
the following subsections.

\subsection{Eigenvalues and eigenfunctions of the angle-dependent equation}

In order to apply NU method [29,30,33,35,36,38-40,42-44], we use a suitable
transformation variable $s=\cos \theta .$ The polar angle part of the Klein
Gordon equation in (26) can be written in the following universal
associated-Legendre differential equation form [38-40]

\begin{equation}
\frac{d^{2}H(s)}{ds^{2}}-\frac{2s}{1-s^{2}}\frac{dH(s)}{ds}+\frac{1}{\left(
1-s^{2}\right) ^{2}}\left[ j(j+D-2)(1-s^{2})-m^{2}-C\alpha _{2}^{2}s^{2}%
\right] H(\theta )=0.
\end{equation}
Equation (27) has already been solved for the three-dimensional
Schr\"{o}dinger equation through the NU method in [38]. However, the aim in
this subsection is to solve with different parameters resulting from the $D$%
-space-dimensions of Klein-Gordon equation. Further, Eq. (27) is compared
with (14) and the following identifications are obtained

\begin{equation}
\widetilde{\tau }(s)=-2s,\text{ \ \ \ }\sigma (s)=1-s^{2},\text{ \ \ }%
\widetilde{\sigma }(s)=-m^{\prime }{}^{2}+(1-s^{2})\nu ^{\prime },
\end{equation}
where
\begin{equation}
\nu ^{\prime }=j^{\prime }(j^{\prime }+D-2)=j(j+D-2)+C\alpha _{2}^{2},\text{
}m^{\prime }{}^{2}=m^{2}+C\alpha _{2}^{2}.
\end{equation}
Inserting the above expressions into equation (22), one obtains the
following function:

\begin{equation}
\pi (s)=\pm \sqrt{(\nu ^{\prime }-k)s^{2}+k-\nu ^{\prime }+m^{\prime }{}^{2}}%
,
\end{equation}
Following the method, the polynomial $\pi (s)$ is found in the following
possible values
\begin{equation}
\pi (s)=\left\{
\begin{array}{cc}
m^{\prime }s & \text{\ for }k_{1}=\nu ^{\prime }-m^{\prime }{}^{2}, \\
-m^{\prime }s & \text{\ for }k_{1}=\nu ^{\prime }-m^{\prime }{}^{2}, \\
m^{\prime } & \text{\ for }k_{2}=\nu ^{\prime }, \\
-m^{\prime } & \text{\ for }k_{2}=\nu ^{\prime }.
\end{array}
\right.
\end{equation}
Imposing the condition $\tau ^{\prime }(s)<0,$ for equation (18), one selects

\begin{equation}
k_{1}=\nu ^{\prime }-m^{\prime }{}^{2}\text{ \ \ and \ \ }\pi (s)=-m^{\prime
}s,
\end{equation}
which yields
\begin{equation}
\tau (s)=-2(1+m^{\prime })s.
\end{equation}
Using equations (19) and (23), the following expressions for $\lambda $ are
obtained, respectively,

\begin{equation}
\lambda =\lambda _{n}=2\widetilde{n}(1+m^{\prime })+\widetilde{n}(\widetilde{%
n}-1),
\end{equation}
\begin{equation}
\lambda =\nu ^{\prime }-m^{\prime }{}(1+m^{\prime }).
\end{equation}
We compare equations (34) and (35)$,$ the new angular momentum\ $j$ values
are obtained as

\begin{equation}
j=-\frac{(D-2)}{2}+\frac{1}{2}\sqrt{(D-2)^{2}+(2\widetilde{n}+2m^{\prime
}+1)^{2}-4C\alpha _{2}^{2}-1},
\end{equation}
or

\begin{equation}
j^{\prime }=-\frac{(D-2)}{2}+\frac{1}{2}\sqrt{(D-2)^{2}+(2\widetilde{n}%
+2m^{\prime }+1)^{2}-1}.
\end{equation}
Using Eqs (15)-(17) and (20)-(21), the polynomial solution of $y_{n}$ is
expressed in terms of Jacobi polynomials [39,40] which are one of the
orthogonal polynomials:

\begin{equation}
H_{\widetilde{n}}(\theta )=N_{\widetilde{n}}\sin ^{m^{\prime }}(\theta )P_{%
\widetilde{n}}^{(m^{\prime },m^{\prime })}(\cos \theta ),
\end{equation}
where $N_{\widetilde{n}}=\frac{1}{2^{m^{\prime }}(\widetilde{n}+m^{\prime })!%
}\sqrt{\frac{(2\widetilde{n}+2m^{\prime }+1)(\widetilde{n}+2m^{\prime })!%
\widetilde{n}!}{2}}$ is the normalization constant determined by $%
\int\limits_{-1}^{+1}\left[ H_{\widetilde{n}}(s)\right] ^{2}ds=1$ and using
the orthogonality relation of Jacobi polynomials [35,45,46]. Besides
\begin{equation}
\widetilde{n}=-\frac{(1+2m^{\prime })}{2}+\frac{1}{2}\sqrt{%
(2j+1)^{2}+4j(D-3)+4C\alpha _{2}^{2}},
\end{equation}
where $m^{\prime }$ is defined by equation (29).

\subsection{Eigensolutions of the radial equation}

The solution of the radial part of Klein-Gordon equation, Eq. (24), for the
Kratzer's potential has already been solved by means of NU-method in [39].
Very recently, using the same method, the problem for the non-central
potential in (2) has been solved in three dimensions ($3D$) by Cheng and Dai
[36]. However, the aim of this subsection is to solve the problem with a
different radial separation function $g(r)$ in any arbitrary dimensions. In
what follows, we present the exact bound-states (real) solution of Eq. (24).
Letting
\begin{equation}
\varepsilon ^{2}=\alpha _{1}^{2}\alpha _{2}^{2},\text{ 4}\gamma
^{2}=(M-1)(M-3)+4B\alpha _{2}^{2},\text{ }\beta ^{2}=A\alpha _{2}^{2},
\end{equation}
and substituting these expressions in equation (24), one gets
\begin{equation}
\frac{d^{2}g(r)}{dr^{2}}+\left( \frac{-\varepsilon ^{2}r^{2}+\beta
^{2}r-\gamma ^{2}}{r^{2}}\right) g(r)=0.
\end{equation}
To apply the conventional NU-method, equation (41) is compared with (14),
resulting in the following expressions:

\begin{equation}
\widetilde{\tau }(r)=0,\text{ \ \ \ }\sigma (r)=r,\text{ \ \ }\widetilde{%
\sigma }(r)=-\varepsilon ^{2}r^{2}+\beta ^{2}r-\gamma ^{2}.
\end{equation}
Substituting the above expressions into equation (22) gives

\begin{equation}
\pi (r)=\frac{1}{2}\pm \frac{1}{2}\sqrt{4\varepsilon ^{2}r^{2}+4(k-\beta
^{2})r+4\gamma ^{2}+1}.
\end{equation}
Therefore, we can determine the constant $k$ by using the condition that the
discriminant of the square root is zero, that is
\begin{equation}
k=\beta ^{2}\pm \varepsilon \sqrt{4\gamma ^{2}+1},\text{ }4\gamma
^{2}+1=(D+2j-2)^{2}+4B\alpha _{2}^{2}.
\end{equation}
In view of that, we arrive at the following four possible functions of $\pi
(r):$%
\begin{equation}
\pi (r)=\left\{
\begin{array}{cc}
\frac{1}{2}+\left[ \varepsilon r+\frac{1}{2}\sqrt{4\gamma ^{2}+1}\right] &
\text{\ for }k_{1}=\beta ^{2}+\varepsilon \sqrt{4\gamma ^{2}+1}, \\
\frac{1}{2}-\left[ \varepsilon r+\frac{1}{2}\sqrt{4\gamma ^{2}+1}\right] &
\text{\ for }k_{1}=\beta ^{2}+\varepsilon \sqrt{4\gamma ^{2}+1}, \\
\frac{1}{2}+\left[ \varepsilon r-\frac{1}{2}\sqrt{4\gamma ^{2}+1}\right] &
\text{\ for }k_{2}=\beta ^{2}-\varepsilon \sqrt{4\gamma ^{2}+1}, \\
\frac{1}{2}-\left[ \varepsilon r-\frac{1}{2}\sqrt{4\gamma ^{2}+1}\right] &
\text{\ for }k_{2}=\beta ^{2}-\varepsilon \sqrt{4\gamma ^{2}+1}.
\end{array}
\right.
\end{equation}
The correct value of $\pi (r)$ is chosen such that the function $\tau (r)$
given by Eq. (18) will have negative derivative [29]. So we can select the
physical values to be

\begin{equation}
k=\beta ^{2}-\varepsilon \sqrt{4\gamma ^{2}+1}\text{ \ \ and \ \ }\pi (r)=%
\frac{1}{2}-\left[ \varepsilon r-\frac{1}{2}\sqrt{4\gamma ^{2}+1}\right] ,
\end{equation}
which yield
\begin{equation}
\tau (r)=-2\varepsilon r+(1+\sqrt{4\gamma ^{2}+1}),\text{ }\tau ^{\prime
}(r)=-2\varepsilon <0.
\end{equation}
Using Eqs (19) and (23), the following expressions for $\lambda $ are
obtained, respectively,

\begin{equation}
\lambda =\lambda _{n}=2n\varepsilon ,\text{ }n=0,1,2,...,
\end{equation}
\begin{equation}
\lambda =\delta ^{2}-\varepsilon (1+\sqrt{4\gamma ^{2}+1}).
\end{equation}
So we can obtain the Klein Gordon energy eigenvalues from the following
relation:
\begin{equation}
\left[ 1+2n+\sqrt{\left( D+2j-2\right) ^{2}+4(\mu +E_{R})B}\right] \sqrt{\mu
-E_{R}}=A\sqrt{\mu +E_{R}},
\end{equation}
and hence for the Kratzer plus the new ring-shaped potential, it becomes

\begin{equation}
\left[ 1+2n+\sqrt{\left( D+2j-2\right) ^{2}+4a_{0}r_{0}^{2}(\mu +E_{R})}%
\right] \sqrt{\mu -E_{R}}=2a_{0}r_{0}\sqrt{\mu +E_{R}},
\end{equation}
with $j$ defined in (36). Although Eq. (51) is exactly solvable for $E_{R}$
but it looks to be little complicated. Further, it is interesting to
investigate the solution for the Coulomb potential. Therefore, applying the
following transformations: $A=Ze^{2},$ $B=0,$ and $j=\ell ,$ the central
part of the potential in (3) turns into the Coulomb potential with Klein
Gordon solution for the energy spectra given by

\begin{equation}
E_{R}=\mu \left( 1-\frac{2q^{2}e^{2}}{q^{2}e^{2}+(2n+2\ell +D-1)^{2}}\right)
,\text{ }n,\ell =0,1,2,...,
\end{equation}
where $q=Ze$ is the charge of the nucleus. Further, Eq. (52) can be expanded
as a series in the nucleus charge as

\begin{equation}
E_{R}=\mu -\frac{2\mu q^{2}e^{2}}{(2n+2\ell +D-1)^{2}}+\frac{2\mu q^{4}e^{4}%
}{(2n+2\ell +D-1)^{4}}-O(qe)^{6},
\end{equation}
The physical meaning of each term in the last equation was given in detail
by Ref. [36]. Besides, the difference from the conventional nonrelativistic
form is because of the choice of the vector $V(r,\theta ,\varphi )$ and
scalar $S(r,\theta ,\varphi )$ parts of the potential in Eq. (1).

Overmore, if the value of $j$ obtained by Eq.(36) is inserted into the
eigenvalues of the radial part of the Klein Gordon equation with the
noncentral potential given by Eq. (51), we finally find the energy
eigenvalues for a bound electron in the presence of a noncentral potential
by Eq. (2) as

\begin{equation}
\left[ 1+2n+\sqrt{\left( 2j^{\prime }+D-2\right)
^{2}+4(a_{0}r_{0}^{2}-C)(\mu +E_{R})}\right] \sqrt{\mu -E_{R}}=2a_{0}r_{0}%
\sqrt{\mu +E_{R}},
\end{equation}
where $m^{\prime }=\sqrt{m^{2}+C(\mu +E_{R})}$ and $\widetilde{n}$ is given
by Eq. (39). On the other hand, the solution of the Schr\"{o}dinger
equation, Eq. (9), for this potential has already been obtained by using the
same method in Ref. [39] and it is in the Coulombic-like form:
\begin{equation}
E_{NR}=-\frac{8\mu a_{0}^{2}r_{0}^{2}}{\left[ 2n+1+\sqrt{(2\ell ^{\prime
}+D-2)^{2}+8\mu (a_{0}r_{0}^{2}-C)}\right] ^{2}},\text{ }n=0,1,2,...
\end{equation}

\begin{equation}
2\ell ^{\prime }+D-2=\sqrt{(D-2)^{2}+\left( 2\widetilde{n}+2m^{\prime
}+1\right) ^{2}-1},
\end{equation}
where $m^{\prime }=\sqrt{m^{2}+2\mu C}.$ Also, applying the following
appropriate transformation: $\mu +E_{R}\rightarrow 2\mu ,$ $\mu
-E_{R}\rightarrow -$ $E_{NR},$ $j\rightarrow \ell $ to Eq. (54) provides
exactly the nonrelativistic limit given by Eq. (55).

In what follows, let us now turn attention to find the radial wavefunctions
for this potential. Substituting the values of $\sigma (r),\pi (r)$ and $%
\tau (r)$ in Eqs (42), (45) and (47) into Eqs. (17) and (21), we find
\begin{equation}
\phi (r)=r^{(\zeta +1)/2}e^{-\varepsilon r},
\end{equation}

\begin{equation}
\rho (r)=r^{\zeta }e^{-2\varepsilon r},
\end{equation}
where $\zeta =\sqrt{4\gamma ^{2}+1}.$ Then from equation (20), we obtain

\begin{equation}
y_{nj}(r)=B_{nj}r^{-\zeta }e^{2\varepsilon r}\frac{d^{n}}{dr^{n}}\left(
r^{n+\zeta }e^{-2\varepsilon r}\right) ,
\end{equation}
and the wave function $g(r)$ can be written in the form of the generalized
Laguerre polynomials as

\begin{equation}
g(\rho )=C_{nj}\left( \frac{\rho }{2\varepsilon }\right) ^{(1+\zeta
)/2}e^{-\rho /2}L_{n}^{\zeta }(\rho ),
\end{equation}
where for Kratzer's potential we have
\begin{equation}
\zeta =\sqrt{\left( D+2j-2\right) ^{2}+4a_{0}r_{0}^{2}(\mu +E_{R})},\text{ }%
\rho =2\varepsilon r.
\end{equation}
Finally, the radial wave functions of the Klein-Gordon equation are obtained
\begin{equation}
R(\rho )=C_{nj}\left( \frac{\rho }{2\varepsilon }\right) ^{(\zeta
+2-D)/2}e^{-\rho /2}L_{n}^{\zeta }(\rho ),
\end{equation}
where $C_{nj}$ is the normalization constant to be determined below. Using
the normalization condition, $\int\limits_{0}^{\infty }R^{2}(r)r^{D-1}dr=1,$
and the orthogonality relation of the generalized Laguerre polynomials, $%
\int\limits_{0}^{\infty }z^{\eta +1}e^{-z}\left[ L_{n}^{\eta }(z)\right]
^{2}dz=\frac{(2n+\eta +1)(n+\eta )!}{n!},$ we have

\begin{equation}
C_{nj}=\left( 2\sqrt{\mu ^{2}-E_{R}^{2}}\right) ^{1+\frac{\zeta }{2}}\sqrt{%
\frac{n!}{\left( 2n+\zeta +1\right) \left( n+\zeta \right) !}}.
\end{equation}
Finally, we may express the normalized total wave functions as

\[
\psi (r,\theta ,\varphi )=\frac{\left( 2\sqrt{\mu ^{2}-E_{R}^{2}}\right) ^{1+%
\frac{\zeta }{2}}}{2^{m^{\prime }}(\widetilde{n}+m^{\prime })!}\sqrt{\frac{(2%
\widetilde{n}+2m^{\prime }+1)(\widetilde{n}+2m^{\prime })!\widetilde{n}!n!}{%
2\pi \left( 2n+\zeta +1\right) \left( n+\zeta \right) !}}
\]
\begin{equation}
\times r^{\frac{(\zeta +2-D)}{2}}\exp (-\sqrt{\mu ^{2}-E_{R}^{2}}%
r)L_{n}^{\zeta }(2\sqrt{\mu ^{2}-E_{R}^{2}}r)\sin ^{m^{\prime }}(\theta
)P_{n}^{(m^{\prime },m^{\prime })}(\cos \theta )\exp (\pm im\varphi ).
\end{equation}
where $\zeta $ is defined in Eq. (61) and $m^{\prime }$ is given after the
Eq. (54).

\section{Conclusions}

\label{C}The relativistic spin-$0$ particle $D$-dimensional Klein-Gordon
equation has been solved easily for its exact bound-states with equal scalar
and vector ring-shaped Kratzer potential through the conventional NU method.
The analytical expressions for the total energy levels and eigenfunctions of
this system can be reduced to their conventional three-dimensional space
form upon setting $D=3.$ Further, the noncentral potentials treated in [30]
can be introduced as perturbation to the Kratzer's potential by adjusting
the strength of the coupling constant $C$ in terms of $a_{0},$ which is the
coupling constant of the Kratzer's potential. Additionally, the radial and
polar angle wave functions of Klein-Gordon equation are found in terms of
Laguerre and Jacobi polynomials, respectively. The method presented in this
paper is general and worth extending to the solution of other interaction
problems. This method is very simple and useful in solving other complicated
systems analytically without given a restiction conditions on the solution
of some quantum systems as the case in the other models. We have seen that
for the nonrelativistic model, the exact energy spectra can be obtained
either by solving the Schr\"{o}dinger equation in (9) (cf. Ref. [39] or Eq.
(55)) or by applying appropriate transformation to the relativistic solution
given by Eq. (54). Finally, we point out that these exact results obtained
for this new proposed form of the potential (2) may have some interesting
applications in the study of different quantum mechanical systems, atomic
and molecular physics.

\acknowledgments This research was partially supported by the
Scientific and Technological Research Council of Turkey. S.M.
Ikhdair wishes to dedicate this work to his family for their love
and assistance.

\end{document}